\documentclass[%
 reprint,
 amsmath,amssymb,
 aps,
]{revtex4-2}
\usepackage{graphicx}
\usepackage{float}
\usepackage{amsmath}

\begin{document}
\title{Kinetically-driven ekpyrosis}
\author{David Shlivko}
\affiliation{Department of Physics, Princeton University, Princeton, NJ 08544, USA}
\email{dshlivko@princeton.edu}
\date{\today}
\begin{abstract}
	We explore the possibility of a scalar field driving ekpyrotic contraction through a non-canonical kinetic energy density rather than a negative potential. We find that this kinetically-driven ekpyrosis (``k-ekpyrosis'') can be achieved in a variety of models, including scalar field theories with power-law, polynomial, or DBI-like kinetic terms in the action. Of these examples, the ekpyrotic phase is best sustained in power-law models, which can generate large and constant equation-of-state parameters, followed by DBI-like models, which can exhibit dynamical attractors toward similarly large equations of state. We show that for a broad class of theories including these examples, phases of k-ekpyrosis are accompanied by preceding or concurrent phases of superluminality.
\end{abstract} 
\maketitle


\section{Introduction} \label{sec:intro}

Ekpyrotic contraction is a dynamical mechanism for producing a homogeneous, isotropic, and spatially flat universe, in which the local Hubble radius shrinks rapidly compared to the decreasing scale factor \cite{bouncing_simple, buchbinder, new_cyclic}. This scenario precedes the Big Bang and can be embedded within the framework of a classical bouncing cosmology, which avoids the geodesic incompleteness, entropy, and quantum runaway (multiverse) problems of inflation \cite{borde, eternal_inflation, new_cyclic}. However, much like inflation, ekpyrotic contraction is impossible to achieve without the existence of a novel form of stress-energy. Even in a universe that is already homogeneous, trace amounts of curvature and anisotropy can be amplified during contraction in accordance with the generalized Friedmann equation \cite{garfinkle},
\begin{equation}\label{friedmann}
	H^2 = \frac{\rho(a)}{3} - \frac{k}{a^2} + \frac{\sigma^2}{a^6},
\end{equation}
where $H$ is the Hubble parameter, $a$ is the scale factor, $\rho$ is the energy density, and $k$ and $\sigma^2$ are the curvature and anisotropy at an initial time corresponding to $a = 1$. Assuming a single form of matter dominates the energy density term, we can use the continuity equation 
\begin{equation}\label{continuity}
	\frac{d(\log \rho)}{d(\log a)} = -2\epsilon(a),
\end{equation}
with equation of state
\begin{equation}\label{equation_of_state}
	\epsilon \equiv \frac{3}{2}\left(1 + \frac{P}{\rho}\right),
\end{equation}
to track the evolution of $\rho(a)$. While the equation of state is constant for vacuum energy ($\epsilon = 0$), dust ($\epsilon = 3/2$), and radiation ($\epsilon = 2$), the stress-energy associated with other sources (such as scalar fields) can have varying $\epsilon(a)$.
From Eqs. (\ref{friedmann}) and (\ref{continuity}), it is evident that the energy density term scales faster than $a^{-6}$ and outpaces the growth of anisotropies during contraction if and only if $\epsilon(a) > 3$. Note that the larger the equation of state is than this critical value, the fewer e-folds of contraction of the scale factor are necessary to make $\Omega_k \equiv \frac{-k}{a^2H^2}$ and $\Omega_\sigma \equiv \frac{\sigma^2}{a^6H^2}$ negligibly small. 


It is straightforward to show that a scalar field $\varphi$ with canonical kinetic energy density $X = -\frac{1}{2} \partial_\mu\varphi \partial^\mu\varphi$ will satisfy the condition $\epsilon > 3$ if and only if it has a negative potential energy density, with large and steep negative potentials allowing large equations of state to be sustained for long periods of time. In particular, a field with a negative exponential potential $V(\varphi) = -V_0 e^{\varphi/m}$ admits stable trajectories with $\epsilon = \frac{1}{2m^2}$; for a sub-Planckian mass scale $m \sim 0.1$, this corresponds to an attractor solution with $\epsilon \sim 50$ \cite{new_cyclic, cook}. Numerical simulations have demonstrated that such a field can not only suppress trace amounts of spatial curvature and anisotropy in a homogeneous universe, but also drive a substantially inhomogeneous, anisotropic, and curved universe toward a flat Friedmann-Robertson-Walker (FRW) fixed point \cite{cook, ultralocality}. 

The goal of the present work is to explore analytically whether similarly large equations of state can be reached and maintained by scalar fields \emph{without} negative potentials but with non-canonical kinetic energy densities that are nonlinear functions of $X$. We refer to these kinetically-driven models as \emph{k-ekpyrosis} in analogy to previous studies of k-inflation \cite{k-inflation, garriga} and k-essence \cite{k-essence, k-essence2} in different contexts. We also note that while negative potentials are useful when building cyclic models of cosmology \cite{new_cyclic}, they are not a necessary component of bouncing models that connect semi-infinite periods of contraction and expansion. 

The general setup of our analysis is described in Sec. \ref{sec:general}. In Sec. \ref{sec:px}, we consider models whose Lagrangians depend only on $X$ and not on the field value, and we show that large and constant equations of state are produced by power-law Lagrangians $\mathcal{L} \propto X^\alpha$ when $\alpha < 1$. In Sec. \ref{sec:polynomial}, we show that field-dependent polynomial Lagrangians of the form $f_n(\varphi)X^n$ can be used to generate ghost-free phases of k-ekpyrosis at third order or higher, but these phases are brief and largely ineffective without fine-tuning of parameters. In contrast, we show in Sec. \ref{sec:dbi} that models with ``wrong-sign'' Dirac-Born-Infeld (DBI) Lagrangians can exhibit stable dynamical attractors at large $\epsilon$ for certain initial conditions. In each of these examples, we discover the appearance of superluminally-propagating field perturbations, whose connection to k-ekpyrosis we discuss in Sec. \ref{sec:superluminal_discussion} in greater generality. We summarize our findings and conclude in Sec. \ref{sec:conclusion}.


\section{General model}\label{sec:general}

Throughout this paper, we will work within the framework of standard General Relativity (GR), with the gravitational sector represented by the Einstein-Hilbert action 
\begin{equation}
	S_g = \frac{1}{2}\int d^4 x \sqrt{-g} R.
\end{equation}
Here, $g$ is the metric determinant, $R$ is the Ricci scalar, and we use reduced Planck units with $c = \hbar = 8\pi G = 1$. For simplicity, we assume the metric of a spatially flat, homogeneous, but anisotropic universe corresponding to the line element
\begin{equation}
	ds^2 = -dt^2 + a(t)^2 e^{2\beta_i(t)}dx_i^2,
\end{equation}
with the constraint $\sum \beta_i = 0$ allowing us to treat $a(t)$ as the (geometric) mean of the three independent scale factors $a_i \equiv a e^{\beta_i}$. In this parameterization, $i \in \{1, 2, 3\}$ labels the spatial dimension, the initial anisotropy appearing in the generalized Friedmann equation (\ref{friedmann}) is $\sigma^2 \equiv \frac{1}{6}\sum \dot{\beta}_i^2|_{a=1}$, and the Hubble parameter is given by $H \equiv \dot{a}/a$, where dots denote derivatives $d/dt$. Note that $H < 0$ in the context of a contracting universe.

We also introduce a scalar field whose action takes the form
\begin{equation}
	S_\varphi = \int d^4x \sqrt{-g} \; P(X, \varphi),
\end{equation}
where the Lagrange density $P$ is allowed to be an arbitrary function of the field value $\varphi$ and the canonical kinetic energy density $X$. Within the scope of this work, we will only be considering models where $P$ increases monotonically with $X$ to ensure that the field's equation of motion 
\begin{equation}
	P_{,\varphi} + \nabla_\mu(P_{,X} \partial^\mu \varphi) = 0
\end{equation}
is everywhere hyperbolic and the emergent geometry is non-singular \cite{bruneton,babichev}. Here and throughout, $\nabla_\mu$ is the covariant derivative and commas represent partial differentiation.

In the homogeneous universe under consideration, the equation of motion reduces to

\begin{equation}\label{eom_general}
	P_{,X} \ddot{\varphi} = P_{,\varphi} - 3HP_{,X}\dot{\varphi} - \dot{P}_{,X}\dot{\varphi},	
\end{equation}
where
\begin{equation}\label{pxd}
	\dot{P}_{,X} =\frac{P_{,XX}(P_{,\varphi}/P_{,X} - 3H\dot{\varphi}) + P_{,X\varphi}}{\dot{\varphi}^{-1} + \dot{\varphi}P_{,XX}/P_{,X}}.
\end{equation}
Additionally, since $\partial_\mu\varphi$ is purely timelike, the field's stress-energy tensor
\begin{equation}\label{T}
	T_{\mu \nu} \equiv \frac{-2}{\sqrt{-g}}\frac{\delta S_\varphi}{\delta g^{\mu\nu}} = P_{,X} \partial_\mu\varphi\partial_\nu\varphi + Pg_{\mu \nu}
\end{equation}
has the same form as that of a perfect fluid, $T_{\mu\nu} = (P + \rho) u_\mu u_\nu  + Pg_{\mu\nu},$
with normalized four-velocity $u_\mu = \partial_\mu\varphi / \sqrt{2X}$, pressure $P$ equal to the field's Lagrange density, and energy density
\begin{equation}\label{rho}
	\rho  = 2X P_{,X} - P.
\end{equation}
The field's equation of state is therefore given by 
\begin{equation}\label{equation_of_state_px}
	\epsilon  = \frac{3XP_{,X}}{2XP_{,X} - P}.
\end{equation}
In the case of a canonical scalar field with $P = X - V(\varphi)$, we have that $P_{,X}$ = 1, and we see that $\epsilon = 3X/(X+V)$ can only satisfy the ekpyrotic condition $\epsilon > 3$ with a negative potential energy density. On the other hand, a non-canonical field can satisfy this condition with a potential of either sign --- or no potential term at all. 

In general, the equation of state of a scalar field may vary over time; however, in this work, we will focus on identifying models in which it is either constant (analogous to the case of $P = X$, where $\epsilon = 3$) or dynamically driven toward a fixed value in the limit of a flat FRW background. An advantage of requiring (asymptotically) constant $\epsilon$ is that one avoids rapid divergences of the equation of state that allow super-Planckian pressures to arise at low energy densities, well before $\rho \sim H^2$ has grown sufficiently to smooth and flatten the universe. (Models with slow divergences such as $\epsilon \sim \log|H|$ may be benign but are beyond the scope of this analysis.) Our approach will be to search for models that admit individual field trajectories with constant $\epsilon$ and then to verify that the equation of state is stable to small deviations from these trajectories in phase space. We will refer to trajectories with $\epsilon = const$ as ``scaling solutions,'' because they correspond to cosmological solutions in which the scale factor evolves as a power law ($a \propto t^{1/\epsilon}$) when the field dominates over other contributions to the Friedmann equation. 

Throughout the examples we consider, we will find that the adiabatic speed of sound
\begin{equation}\label{cs2}
	c_s^2 \equiv \frac{P_{,X}}{\rho_{,X}} = \frac{P_{,X}}{P_{,X} + 2XP_{,XX}},
\end{equation}
which is equal to unity for canonical fields with $P_{,XX} = 0$, often exceeds the speed of light in models of kinetically-driven ekpyrosis. This sound speed characterizes the propagation of both scalar field perturbations on a homogeneous background \cite{perturbations_eom} and scalar perturbations to the gravitational metric \cite{garriga}. A discussion of whether these superluminal dynamics are physical is deferred to Sec. \ref{sec:superluminal_discussion}.

\begin{figure*}[t]
 \begin{center}
	\includegraphics[width=\textwidth]{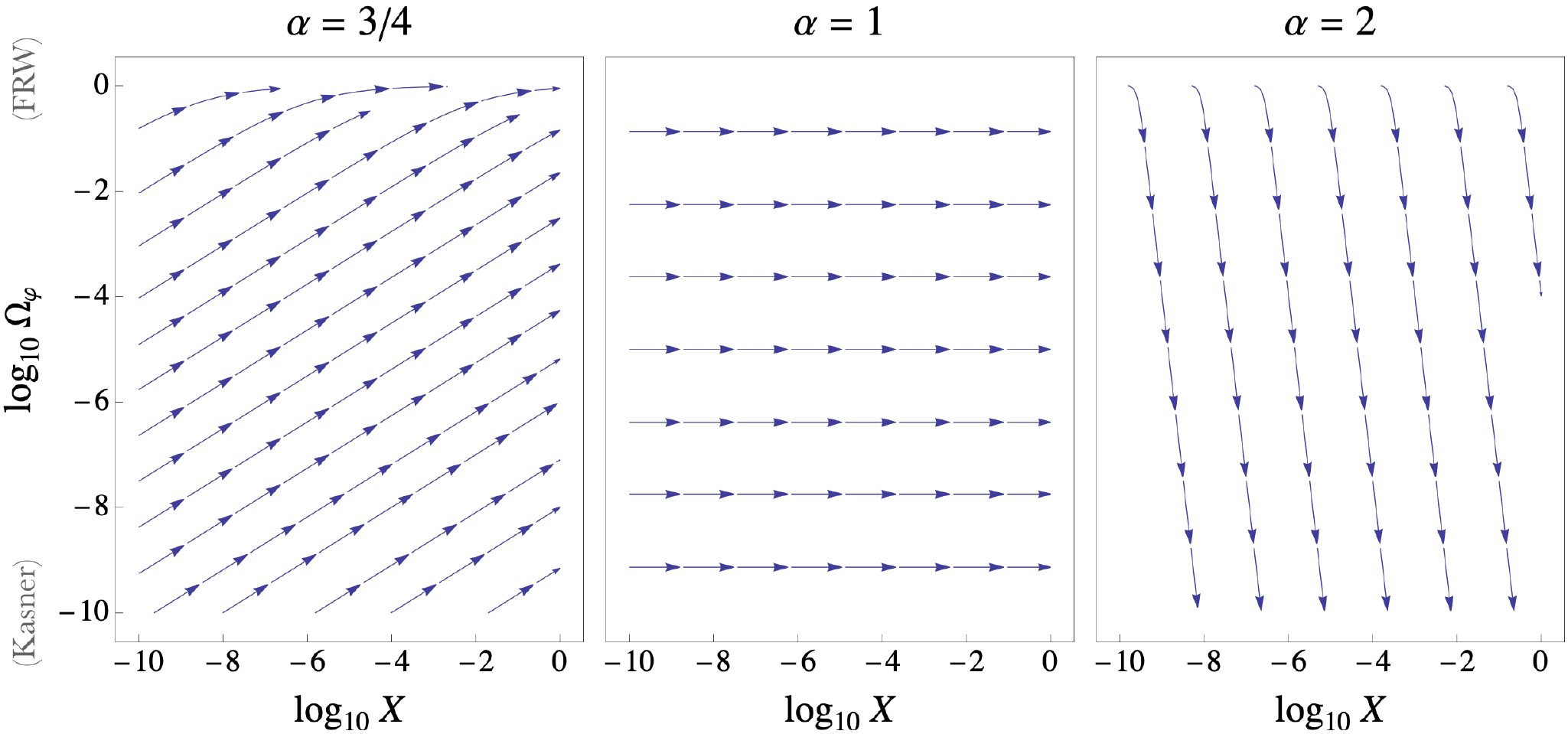}
	\caption{\label{fig_stream_power} Evolution in phase space of the relative energy density parameter $\Omega_\varphi = 1 - \Omega_\sigma$ and the canonical kinetic term $X = \frac{1}{2} \dot{\varphi}^2$ in a homogeneous and spatially flat but anisotropic universe. Arrows indicate the flow as the universe contracts, and each panel represents a different power-law model $P(X) \propto X^\alpha$. The left panel is representative of models with $\alpha < 1$, in which the universe is dynamically driven toward isotropy ($\Omega_\varphi \approx 1$) through k-ekpyrosis. The middle panel illustrates the Kasner-like evolution produced by the canonical model with $\alpha = 1$, and the right panel represents models with $\alpha > 1$ that drive the universe toward pure Kasner spacetime ($\Omega_\sigma \approx 1$).
}
\end{center}
\end{figure*}

\section{P(X) models}\label{sec:px}
In this section, we consider models in which the Lagrange density is independent of the field value, i.e., $P(X, \varphi) = P(X)$. The field's equation of motion (\ref{eom_general}) simplifies to 
\begin{equation}\label{eom_px}
	\ddot{\varphi} = -3H\dot{\varphi}c_s^2,
\end{equation}
where $c_s^2$ is given in Eq. (\ref{cs2}). We see that the field experiences a Hubble anti-friction effect that repels it from the vacuum solution $\dot{\varphi} = 0$ in a contracting universe ($H < 0$). 

Our goal in this section is to search for specific $P(X)$ theories that admit scaling solutions on a flat FRW background with constant $\epsilon > 3$, as motivated in Secs. \ref{sec:intro}-\ref{sec:general}. Because any nonzero $X$ increases monotonically during contraction, the condition $\dot{\epsilon} = 0$ is equivalent to the requirement that $\epsilon$ is independent of $X$. Assuming $P_{,X}$ is nowhere vanishing, the relation (\ref{equation_of_state_px}) can be rewritten as
\begin{equation}
	\frac{d(\log P)}{d(\log X)} = \frac{1}{2-3/\epsilon}
\end{equation}
and solved under the condition of constant $\epsilon$ to find that
\begin{equation}\label{power_law}
	P \propto X^{\alpha},
\end{equation}
where the power $\alpha$ is related to $\epsilon$ via
\begin{equation}
	\epsilon = \frac{3\alpha}{2\alpha-1}.
\end{equation}
We note that models in which $\epsilon$ depends on $X$ but asymptotically approaches a fixed value will simply correspond to Lagrangians $P(X)$ that asymptotically approach the form (\ref{power_law}).

The range of powers $\alpha \in (1/2, \; 1)$ corresponds to $\epsilon > 3$, with limiting behavior $\epsilon \to \infty$ as $\alpha \to 1/2$. We note that this ekpyrotic parameter space coincides with the regime where the adiabatic sound speed is superluminal, with
\begin{equation}
	c_s^2 = \frac{1}{2\alpha-1}.
\end{equation}

On the other hand, any model with $\alpha > 1$ will have $\epsilon < 3$, and the choice $\alpha = 1$ reproduces the canonical free field theory with $P(X) = X$ and $\epsilon = 3$. As we have shown in Sec. \ref{sec:intro}, only the models with $\epsilon > 3$ will reduce the relative anisotropy $\Omega_\sigma$, while the models with $\epsilon < 3$ will instead cause the universe to tend toward a pure Kasner state ($\Omega_\sigma = 1$). In the intermediate case ($\epsilon = 3$), the field's energy density and the anisotropy term scale synchronously as $a^{-6}$, causing the relative anisotropy to remain constant in what has been called a ``Kasner-like'' state \cite{robustness, ultralocality}. These three scenarios are compared in Fig. \ref{fig_stream_power}, where we plot the phase-space evolution of $X$ (which generically increases over time) and the energy density parameter $\Omega_\varphi = 1 - \Omega_\sigma$. 
In the ekpyrotic model with $\alpha < 1$ (depicted in the left panel), we see that every e-fold of $X$ (and correspondingly of $|H|$)  produces a consistent number of e-folds of $\Omega_\varphi$ until it nears the asymptote of $\Omega_\varphi = 1$. 

In the field-dominated limit ($\Omega_\varphi \approx 1$), the trajectory and scaling relations for a model with $\alpha < 1$ become
\begin{equation}
		H(t) = -\frac{1}{\epsilon|t|}, \quad a(t) \propto |t|^{1/\epsilon}, \quad \varphi(t) \propto |t|^{3/\epsilon - 1}.
\end{equation}
Here, the field is defined up to an arbitrary additive constant, and the time coordinate is chosen such that physical quantities diverge at $t = 0$. (In a classical bouncing cosmology, these divergences are avoided due to a transition from contraction to expansion at some earlier time.) In a representative model with $\alpha = 3/4$ (corresponding to $\epsilon = 4.5$), we see that every e-fold of $\varphi$ corresponds to 3 e-folds of contraction of the Hubble radius $|H|^{-1}$ and a similarly large suppression of $\Omega_k \propto |H|^{-14/9}$ and $\Omega_\sigma \propto |H|^{-2/3}$.

\section{Field-coupled polynomial models}\label{sec:polynomial}
In previous investigations of k-inflation \cite{k-inflation} and k-essence \cite{k-essence}, the authors considered a simple quadratic Lagrangian of the form 
\begin{equation}\label{quadratic}
	P(X, \varphi) = K(\varphi)X + L(\varphi)X^2,
\end{equation}
with a ``wrong-sign'' kinetic term $K < 0$ but a positive quartic term $L > 0$. This setup ensures that there is a positive energy cost
\begin{equation}
	\rho = K(\varphi)X + 3L(\varphi)X^2
\end{equation}
 to having large $|X|$, while simultaneously allowing for the existence of negative pressures at low $X$. To achieve the opposite effect ($P \gg \rho$) in a regime where $\rho > 0$, one must instead have $K > 0$ and $L < 0$, which leads to a severe gradient instability in addition to violating the condition $P_{,X} > 0$ for global hyperbolicity. 
For completeness, we note that these unstable models exhibit the same correspondence observed in Sec. \ref{sec:px} between ekpyrosis ($\epsilon > 3$) and superluminality ($c_s^2 > 1$) in regions of phase space where $c_s^2 > 0$.

If we extend our analysis to cubic models of the form
\begin{equation}
	P(X, \varphi) = K(\varphi)X + L(\varphi)X^2 + Q(\varphi)X^3,
\end{equation}
it becomes possible to choose functions $K > 0, L < 0, Q > 0$ in such a way that $P_{,X}$ is everywhere positive but has a nontrivial structure that allows the equation of state
\begin{equation}
	\epsilon = 3\, \frac{KX + 2LX^2 + 3QX^3}{KX + 3LX^2 + 5QX^3}
\end{equation}
to exceed $3$ in some regions of phase space. A simple example is the Lagrange density
\begin{equation}\label{cubic_example}
	P(X, \varphi) = X - \varphi^2 X^2 + \varphi^4 X^3,
\end{equation}
which corresponds to the energy density
\begin{equation}
	\rho(X, \varphi) = X - 3\varphi^2 X^2 + 5\varphi^4 X^3.
\end{equation}
\begin{figure}
	\includegraphics[width=\columnwidth]{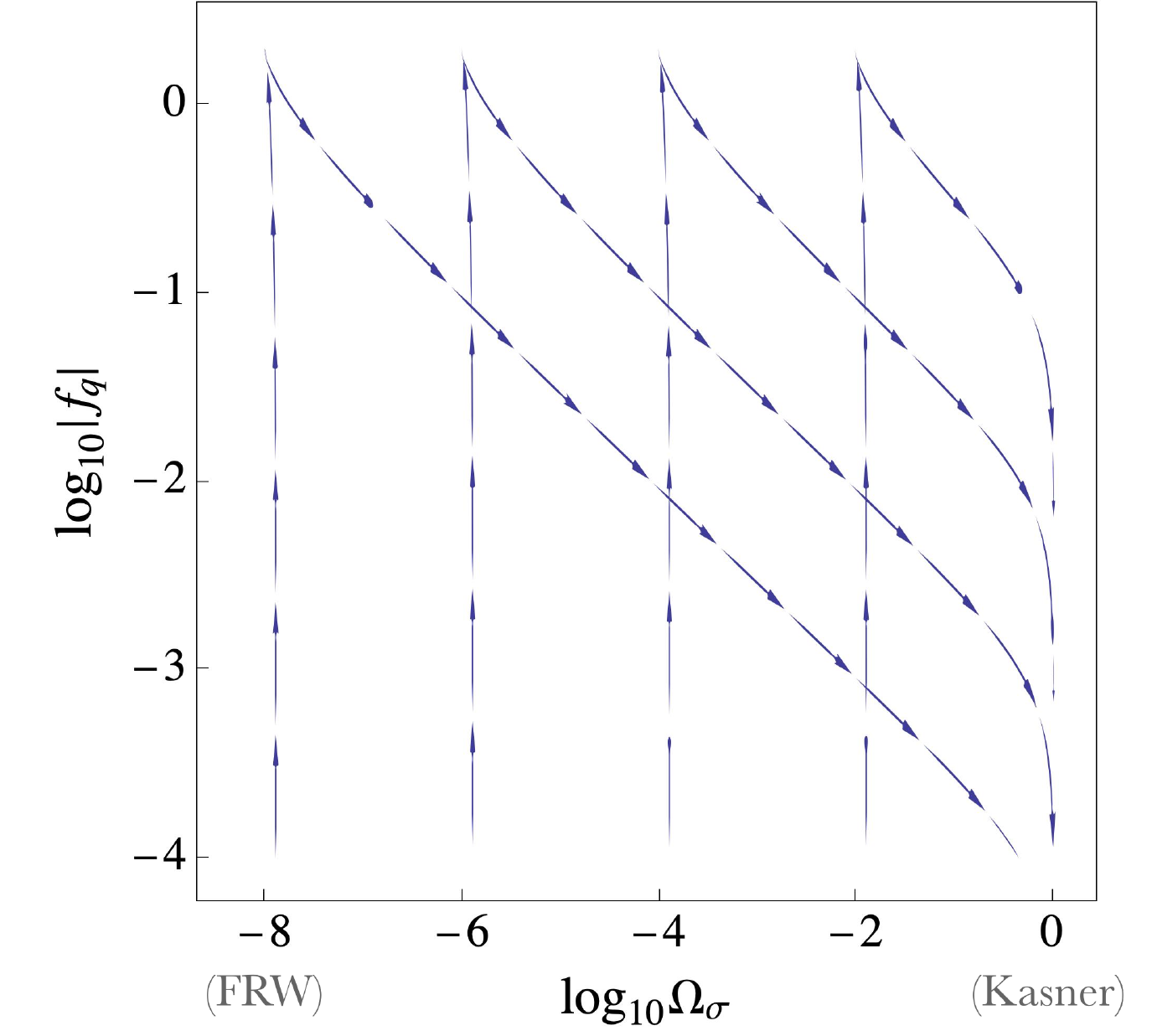}
	\caption{\label{fig_stream_cubic} Evolution in phase space of the anisotropy parameter $\Omega_\sigma$ and the relative contribution of the quadratic energy density term $f_q \equiv \frac{-3\varphi^2 X^2}{\rho}$ for the cubic model given in Eq. (\ref{cubic_example}). The evolution equations are coupled to the Hubble parameter, but its monotonic increase in magnitude is suppressed in the figure for clarity. The trajectories consist of two distinct phases.
	First, in the low-$X$ limit, the linear term is dominant but the contribution $f_q$ of the quadratic term is growing in magnitude. This phase roughly coincides with the period of ekpyrosis where $3 \lesssim \epsilon \lesssim 3.67$, but there is no noticeable suppression of anisotropy by the time trajectories reach the maximum value of $|f_q| \approx 2$ and the first phase concludes. The second phase of the trajectories sees $f_q$ recede toward zero as the cubic term grows dominant over linear and quadratic contributions to the energy density. This phase roughly corresponds to non-ekpyrotic contraction with $1.8 \lesssim \epsilon \lesssim 3$ and leads to a substantial growth in anisotropy. \\}
\end{figure}
 \begin{figure*}[t]
 \begin{center}
	\includegraphics[width=\textwidth]{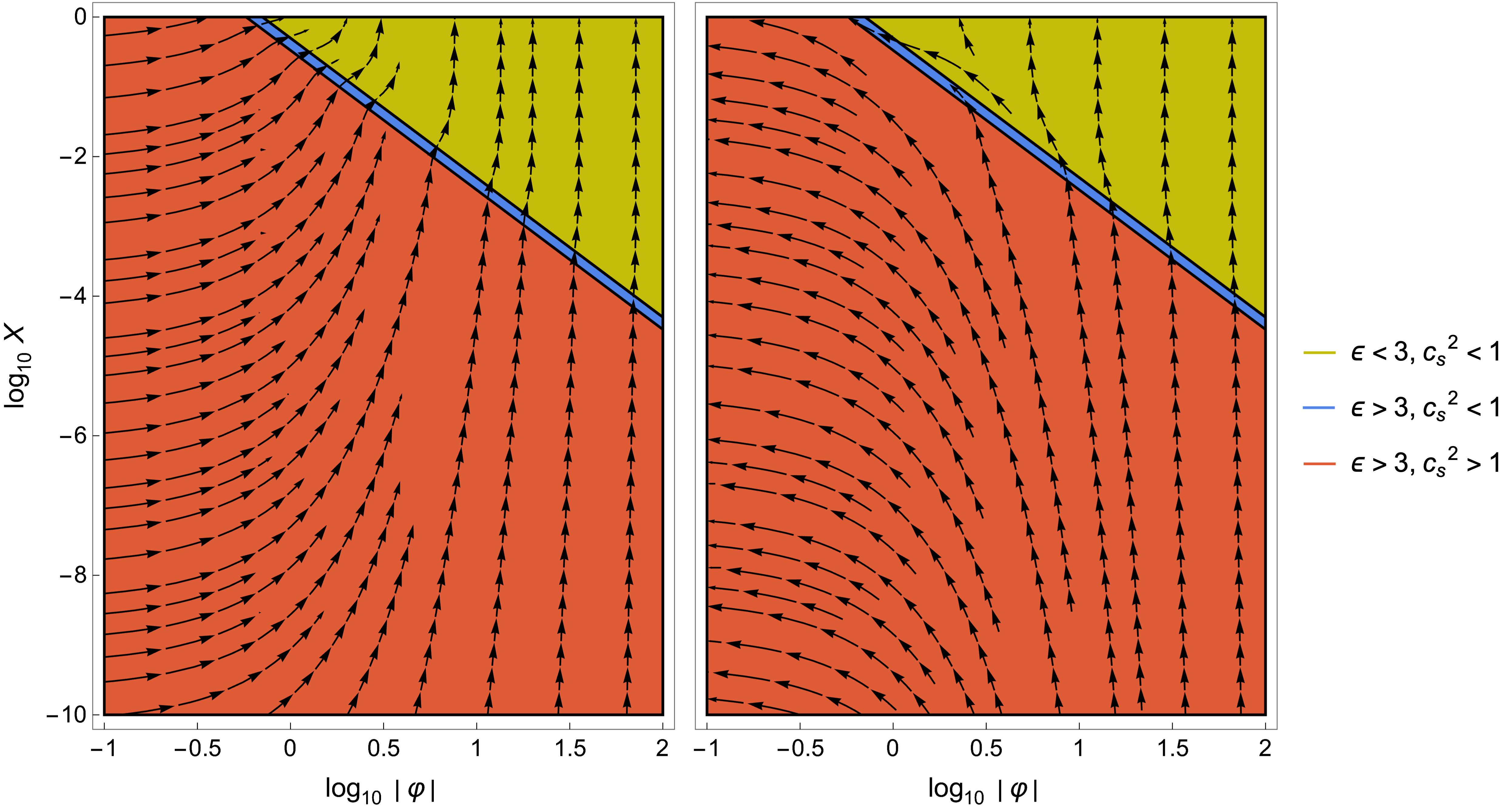}
	\caption{\label{fig_shaded_cubic} The phase diagram of the cubic model (\ref{cubic_example}) depicts a flow of trajectories in an isotropic contracting universe from a superluminal ekpyrotic phase (bottom/red region) to a shorter-lived subluminal ekpyrotic phase (middle/blue strip) to a subluminal non-ekpyrotic phase (top/yellow region). The two panels distinguish between cases where $\varphi$ and $\dot{\varphi}$ have the same sign (left) or different sign (right). In either case, no trajectory can remain fully subluminal unless $X$ is manually bounded from below in the past and $|\varphi|$ is chosen to be sufficiently large.
		}
\end{center}
\end{figure*}

This model reduces to a canonical potential-free scalar field in the low-$X$ limit and is generally well-behaved at any finite $X$ and $\varphi$. Despite being physically viable, however, the model is dynamically limited: the maximum attainable equation of state is $\epsilon \approx 3.67$, and that value is only generated in a ``sweet spot'' between the extremes of $\varphi^2 X \ll 1$ (where the linear term dominates and $\epsilon \approx 3$) and  $\varphi^2 X \gg 1$ (where the cubic term dominates and $\epsilon \approx 1.8$). The relative contribution of the quadratic term to the energy density,
\begin{equation}
	f_q \equiv \frac{-3\varphi^2 X^2}{\rho},
\end{equation}
can be used to track the transition between these extremes. We see in Fig. \ref{fig_stream_cubic} that the weakly ekpyrotic phase with $3 \lesssim \epsilon \lesssim 3.67$, which roughly corresponds to the phase of increasing $|f_q|$, has hardly any effect on $\Omega_\sigma$ before the cubic term becomes relevant, drives $|f_q|$ back down toward zero, and allows anisotropies to grow. 

The cubic model serves an important pedagogical purpose, however, by illustrating that a phase of k-ekpyrosis is not always associated with a simultaneous phase of superluminality. Indeed, the parameters $K$, $L$, and $Q$ allow the sound speed to vary independently of the equation of state. However, we emphasize that it is still not feasible for an entire \emph{trajectory} to remain subluminal while smoothing the universe through k-ekpyrosis. We show in Fig. \ref{fig_shaded_cubic} that the thin strip of phase space admitting subluminal ekpyrosis is bordered by a wider superluminal region at lower $X$. Sample trajectories for the case of an isotropic universe are superposed on these phase space diagrams; they pierce through this strip but spend relatively little time there. In order for ekpyrotic trajectories to avoid the superluminal regime, one would need to impose a \emph{lower bound} $X_\text{min}(\varphi)$ on the domain of validity of the cubic EFT, tuned to lie precisely within the thin strip of Fig. \ref{fig_shaded_cubic}. Moreover, one can see from the figure that the bound at $|\varphi| \lesssim 1$ would need to be super-Planckian, leaving only trajectories with exceptionally large $|\varphi|$ within the scope of the model. The key takeaway is that without such unnatural modifications, superluminality appears generically either before or during a given phase of k-ekpyrosis; we will show in Sec. \ref{sec:superluminal_discussion} that this conclusion generalizes to a wide class of $P(X, \varphi)$ models.

\section{DBI models}\label{sec:dbi}

The divergence of $\epsilon$ in models with $P(X) \propto X^{1/2}$ suggests that the DBI action with
\begin{equation}\label{p_dbi}
	P(X, \varphi) = \frac{-1}{f(\varphi)} \; \left(\sqrt{1-2Xf(\varphi)}-1\right)
\end{equation}
may offer a powerful smoothing mechanism in the limit $-2fX \gg 1$, assuming we allow the warp factor $f(\varphi)$ to be negative. A negative warp factor may seem unnatural, since the DBI action originated as an analog to the classical relativistic kinetic energy $\mathcal{L}_\text{kin} = -mc^2(\sqrt{1-v^2c^{-2}} - 1)$ \cite{born_infeld} and has more recently seen applications within string theory (from which the terminology of warp factors originates) using the original (positive) sign. Nevertheless, the ``wrong-sign'' action maintains the theory's symmetries and has been argued to be free of pathologies as an effective field theory (EFT) \cite{burrage}. DBI models with negative warp factors (which we have referred to as ``DBI-like'' in previous sections) have also been considered in the contexts of cosmological screening mechanisms \cite{burrage, panpanich_2017, panpanich_2019}, kinetic inflation \cite{mukhanov_2006}, and black hole information \cite{black_hole}.

We classify the DBI model as a kinetic theory because the field's pressure and energy density vanish in the limit $X \rightarrow 0$.
To find the DBI field's equation of state, we may use Eq. (\ref{rho}) to obtain
\begin{equation} \label{rho_dbi}
	\rho = \frac{2X}{\sqrt{1-2fX}} - P,
\end{equation}
which then gives
\begin{equation}\label{epsilon_dbi}
	\epsilon = \frac{3}{2}\left(1 + \sqrt{1-2fX}\right).
\end{equation}
As we had expected, ekpyrotic contraction can only be achieved when $f(\varphi) < 0$, and the equation of state is greatest when $X$ is large compared to $f^{-1}$. 
We note here that allowing a negative warp factor also leads the DBI field to exhibit superluminal perturbations, with the sound speed
\begin{equation}
	c_s = \sqrt{1 - 2fX}
\end{equation}
scaling linearly with the equation of state (\ref{epsilon_dbi}). Finally, the field's equation of motion is
\begin{equation} \label{eom_dbi}
	\ddot{\varphi} + \frac{3}{2}\frac{f_\varphi}{f}\dot{\varphi}^2 - \frac{f_\varphi}{f^2} + 3H\dot{\varphi}(1-2fX) + \frac{f_\varphi}{f^2}(1-2fX)^{3/2} = 0,
\end{equation}
where the final term on the left-hand-side is due to the $1/f$ term in the action.

\subsection{DBI scaling solutions}\label{sec:dbi_scaling}

In this section, we will search for scaling solutions in the FRW limit that satisfy $\dot{\epsilon} = 0$ at large $\epsilon$. To proceed, we make the choice
\begin{equation}\label{f_exp}
	f(\varphi) \equiv \alpha e^{2\varphi/m},
\end{equation}
where
\begin{equation}
	\alpha \equiv \frac{8(2-3m^2)}{m^2(4-3m^2)^2} < 0
\end{equation}
and $m$ is a mass scale near order unity in reduced Planck units.  We are free to choose the coefficient $\alpha$ (which serves to simplify later arithmetic) without loss of generality by an appropriate redefinition of the field, $\varphi \mapsto \varphi + \varphi_0$. 

A negative exponential warp factor allows the DBI model to achieve $\epsilon \gg 3$ even at low $X$ and to maintain this equation of state indefinitely as $X$ grows from gravitational blueshifting if the field rolls toward smaller $|f(\varphi)|$. Indeed, unlike other common choices for $f^{-1}$ (e.g., constant \cite{burrage, panpanich_2019, mukhanov_2006, black_hole}, quadratic \cite{kar}, or quartic \cite{wei}), models with an exponential warp factor (used also in \cite{panpanich_2017}) may admit scaling solutions at arbitrarily large $\epsilon$. We show in Appendix \ref{app:scaling_dbi} that these models can have up to two scaling solutions, namely the ``canonical'' solution,
\begin{equation}
	\epsilon_c = 3,
\end{equation}
which corresponds to the formal limit $X \to 0$, and a ``non-canonical'' solution,
 \begin{equation}\label{epsilon_nc}
 	\epsilon_{nc} = \frac{6}{4-3m^2},
 \end{equation}
which is generated by non-linearities in the action and exists only if $m^2 \in (2/3, \,4/3)$ and the field has negative velocity ($\dot{\varphi} < 0$). We include in the appendix a dynamical systems analysis demonstrating that $\epsilon_c$ is a repeller fixed point of the equation of state, while $\epsilon_{nc}$ is an attractor. The result is that trajectories with $\dot{\varphi} < 0$ will be driven asymptotically toward $\epsilon_{nc}$, while trajectories with $\dot{\varphi} > 0$ will see the field's equation of state increase without bound as the universe contracts. Note that the field velocity can never change sign, since the equation of motion (\ref{eom_dbi}) yields $\ddot{\varphi} = 0$ whenever $\dot{\varphi} = 0$.
 
If the DBI field has negative velocity and reaches the $\epsilon_{nc}$ scaling solution before its energy density is large enough to trigger a cosmological bounce,
 we can track its trajectory analytically through the remainder of the contraction phase. In particular, assuming the equation of state (\ref{epsilon_nc}), the equation of motion for the DBI field reduces to 
 \begin{equation} \label{eom_nc}
	\ddot{\varphi} = \frac{4}{m(4-3m^2)} \alpha^{-1} e^{-2\varphi/m}(\sqrt{1 - \alpha e^{2\varphi/m} \dot{\varphi}^2} - 1),
\end{equation}
which has a solution (as long as $m^2 < 4/3$) given by
\begin{equation} \label{traj}
	\varphi(t) = m \log|t|.
\end{equation}
The two degrees of freedom coming from integration constants have been fixed by the requirement that $\epsilon = \epsilon_{nc}$ and by our choice of time coordinate, which is defined such that the field, its energy density, and other physical quantities diverge at $t = 0$.

One finds that a field following the trajectory (\ref{traj}) will have fixed values for
\begin{equation}
	2fX = \alpha m^2, \quad c_s = \sqrt{1-\alpha m^2}, \quad \epsilon = \frac{6}{4-3m^2}
\end{equation}
and analytic solutions for
\begin{equation}\label{analytic}
	f(t) = \alpha t^2, \quad H(t) = -\frac{1}{\epsilon|t|}, \quad a(t) \propto |t|^{1/\epsilon}.
\end{equation}
From the scaling relation for $H(t)$, one can conclude that the DBI field must traverse many Planck masses in field space in order to achieve significant suppression of $\Omega_k$ and $\Omega_\sigma$ due to its logarithmic trajectory and the requirement of an order-unity mass scale $m$.


 \subsection{DBI trajectories with anisotropy}\label{sec:subdominant}
 \begin{figure*}[t!]
 \begin{center}
	\includegraphics[width=0.45\textwidth]{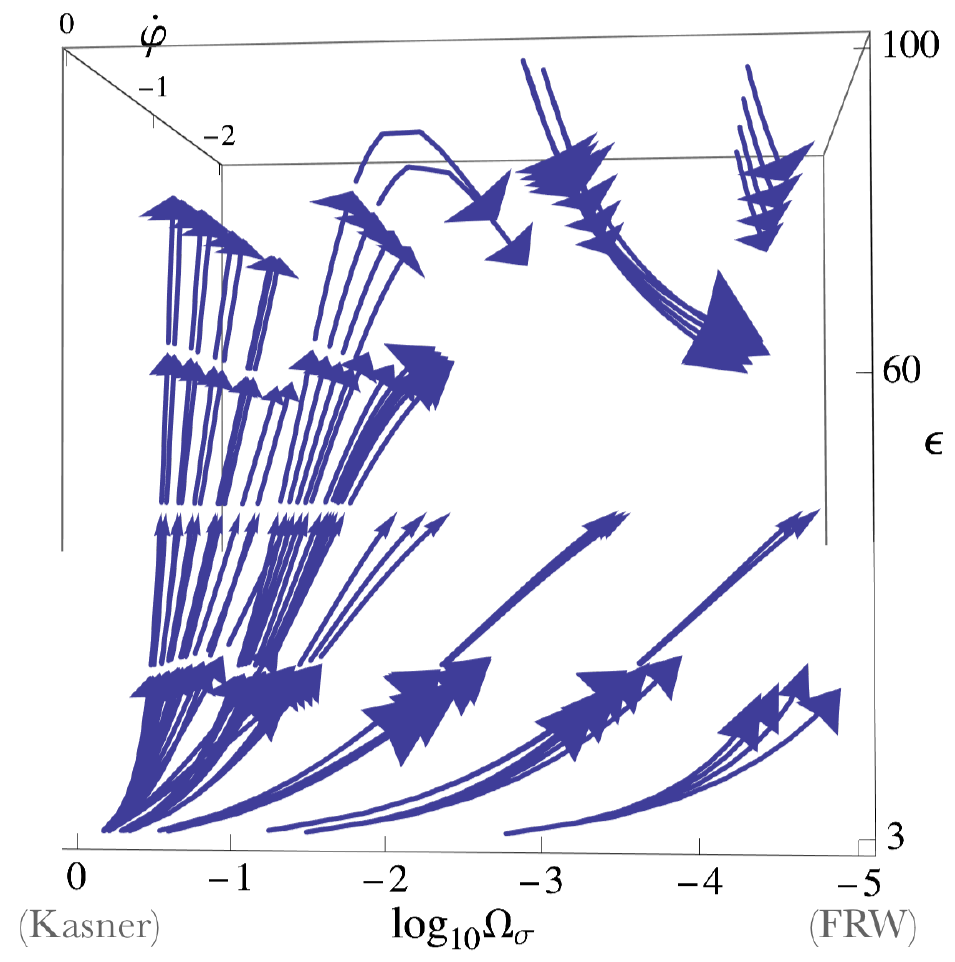}\quad \-\
	\includegraphics[width=0.46\textwidth]{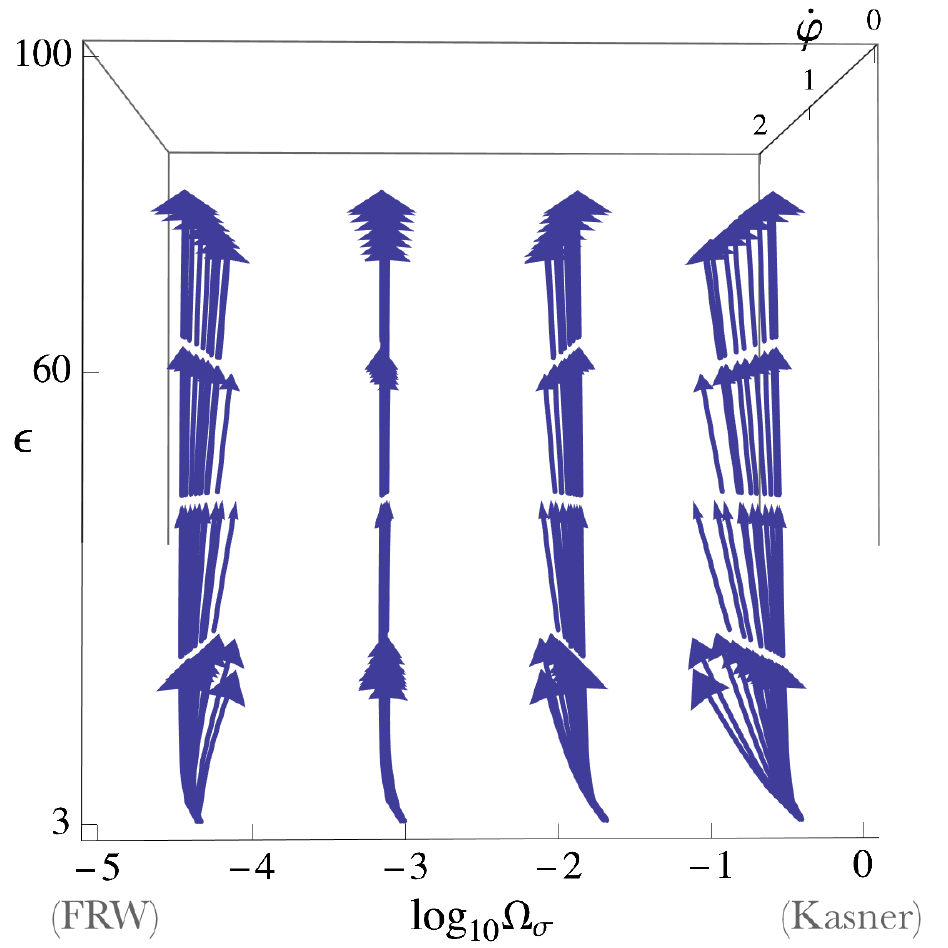}	
	\caption{\label{fig_dbi} Phase space evolution of the equation of state $\epsilon$, the anisotropy parameter $\Omega_\sigma$, and the field velocity $\dot{\varphi}$ for the DBI model defined by Eqs. (\ref{p_dbi}) and (\ref{f_exp}). Qualitatively different behaviors are observed when the field velocity is negative (left panel) versus positive (right panel). If the field rolls with negative velocity, its equation of state increases until it is sufficiently dominant [cf. Eq. (\ref{omega_bound})] for the scaling solution at $\epsilon_{nc} = 60$ to exist and act as a dynamical attractor. On the other hand, if the field rolls with positive velocity, there is no such attractor solution and its equation of state will increase indefinitely. In this latter case, $\epsilon$ can increase by many orders of magnitude well before the resulting ekpyrotic contraction has an observable effect on $\Omega_\sigma$.
	}
\end{center}
\end{figure*}

In this section, we present numerical simulations of DBI trajectories beginning from an anisotropy-dominated epoch of the contracting universe. For simplicity, we neglect the presence of spatial curvature and other forms of stress-energy whose contributions to the Friedmann equation grow more slowly than the anisotropy term. In these simulations, the characteristic mass scale will be set to $m^2 = 39/30$, which corresponds to $\alpha \approx -1000$, $\epsilon_{nc} = 60$ and $c_s = 39$.

 We first consider the scenario where initial conditions set $\dot{\varphi} < 0$, in which case we expect the DBI field's equation of state to be drawn to the $\epsilon_{nc}$ attractor  and ultimately follow the trajectory (\ref{traj}). The scaling solution itself, however, only exists when the DBI field dominates over anisotropy in the Friedmann equation; we show quantitatively in Appendix \ref{app:scaling_sub} that the earliest it can appear is when $\Omega_\varphi \equiv \rho_\varphi/(3H^2) = 3m^2/4$. This behavior is an important feature of the theory, not a bug: if the $\epsilon_{nc}$ fixed point were to persist throughout the semi-infinite contraction phase, then a region of space where the field's equation of state is $\epsilon > \epsilon_{nc}$ (set by an initial condition or quantum fluctuation) would exhibit a divergence in $\epsilon$ when extrapolated backwards in time. Instead, we see in the left panel of Fig. \ref{fig_dbi} that all DBI trajectories ``start out'' at $\epsilon \approx 3$ and are generically repelled from the unstable fixed point at $\epsilon_c = 3$ toward larger equations of state. The trajectories then either asymptotically approach $\epsilon_{nc}$ from below or surpass $\epsilon_{nc}$ during the sub-dominant regime and later approach $\epsilon_{nc}$ from above. 
 
 The situation is somewhat different for positive-velocity trajectories, which lack the $\epsilon_{nc}$ scaling solution altogether. In both the dominant and sub-dominant regimes, the field's equation of state is repelled away from $\epsilon_c$ toward arbitrarily large values (see right panel of Fig. \ref{fig_dbi}). This rapid divergence of $\epsilon$ enables the pressure of the field to reach super-Planckian values even at low energy densities, and in DBI models, it simultaneously leads to unbounded growth of the speed of sound. As a result, a practicable DBI theory would need to incorporate either a taming mechanism for the field's pressure [e.g., by modifying $f(\varphi)$] or a mechanism that disfavors positive-velocity trajectories (e.g., by including an external potential) in order to prevent these issues from arising. 

\section{Superluminality and physical consistency}\label{sec:superluminal_discussion}
Throughout the previous sections, we have seen strong hints of a correlation between ekpyrosis ($\epsilon > 3$) and superluminality ($c_s^2 > 1$) in kinetically-driven models. We saw specifically that models using power-law, quadratic, or DBI-like Lagrangians \emph{required} superluminality to achieve ekpyrosis, while cubic models generated ekpyrotic trajectories that were all mostly (but not entirely) superluminal. Here, we will show more generally that any globally hyperbolic $P(X, \varphi)$ model with a positive definite derivative $P_{,X} > 0$ and a positive semi-definite effective potential $V_\text{eff}(\varphi) \equiv -P(X=0, \varphi) \geq 0$ can only have $\epsilon > 3$ in some domain of phase space if it also has $c_s^2 > 1$ in some (possibly different) domain. To see this, we begin by noting that the condition for ekpyrotic contraction corresponds to having $P > \rho$, or equivalently
\begin{equation}\label{ekpyrotic_contraction}
	\epsilon > 3 \iff P > X P_{,X}.
\end{equation}
At the same time, the speed of sound (\ref{cs2}) is superluminal under the condition
\begin{equation}\label{superluminality}
	c_s^2 > 1 \iff P_{,XX} < 0,
\end{equation}
assuming a homogeneous background with $X > 0$ and a globally hyperbolic theory with $P_{,X} > 0$.
In models with a positive semi-definite potential, the ekpyrotic condition (\ref{ekpyrotic_contraction}) can be satisfied at some $(X, \varphi) = (X^*, \varphi^*)$ only if there is a domain $\mathcal{D} \subset [0, X^*]$ in which $P(X, \varphi^*)$ is concave and therefore satisfies the superluminality condition (\ref{superluminality}). A corollary to this result is that trajectories that are ekpyrotic at arbitrarily small $X$ must simultaneously be superluminal in the limit $X \to 0$. Other ekpyrotic trajectories, which either avoid the small-$X$ limit or have $\epsilon < 3$ in that regime, are still likely to run through a superluminal region of phase space without substantial tuning of the model or initial conditions. 

With this in mind, it is natural to ask whether superluminal models of k-ekpyrosis can be valid field theories (or effective field theories) in the real world. It has previously been shown that typical issues arising in tachyonic theories, such as closed causal curves and their resulting causal paradoxes, \emph{do not} arise for superluminal field perturbations in the cosmological context \cite{babichev}. In particular, it is important to note that the homogeneous cosmological background is a Lorentz-violating medium.
As a consequence, observers in the background frame always see signals sent through this medium propagate forward in time along the sound cone, while observers in boosted frames who may see signals propagate backward in time will notice that a ``return signal'' sent forward in time would travel more slowly, even if it were sent through a different background field with a greater sound speed. It has also been shown that the Cauchy problem remains well-posed in the presence of superluminal fields as long as initial conditions are specified on Cauchy slices that are spacelike with respect to both the light cone and the wider sound cone of the scalar field \cite{babichev, bruneton}. The issue of whether superluminal EFT's are compatible with standard UV completions is beyond the scope of this discussion, but for references see \cite{adams, aoki, crem1, galileon, crem2}.

\section{Conclusions}\label{sec:conclusion}

In this work, we identified and investigated three classes of non-canonical kinetic scalar field theories in which ekpyrotic contraction can occur. These include power-law models of the form $P \propto X^\alpha$, polynomial models of the form $f_n(\varphi)X^n$, and DBI-like models with negative warp factors. The power-law models describe fields with a constant and arbitrarily large equation of state that can drive ekpyrosis ($\epsilon > 3$) when $\alpha \in (1/2, 1)$, and we showed that any theory with a field-independent Lagrangian $P(X)$ can only drive k-ekpyrosis with asymptotically constant $\epsilon$ if $P(X)$ itself asymptotically approaches the form $X^\alpha$.
Fields with a polynomial action can have an ekpyrotic equation of state even at second order, but well-behaved models with positive definite $P_{,X}$ are possible only at third order or higher, and even those models have difficulty achieving and sustaining large $\epsilon$ without finely tuned coefficients $f_n(\varphi)$. 
The robustness of DBI-like models falls between these extremes: the DBI field does not have a fixed equation of state, but it can exhibit a dynamical attractor at large $\epsilon$ when the warp factor is chosen to be exponential in the field value, as long as the characteristic mass scale lies in the range $m^2 \in (2/3, 4/3)$ and initial conditions set $\dot{\varphi} < 0$.

While these examples demonstrate that kinetically-driven ekpyrotic contraction is possible, we have also found that theories of k-ekpyrosis generally come with subtleties and complications that are absent in the canonical, potential-driven scenario. Most notably, we saw that scalar perturbations to the field and gravitational metric can propagate superluminally in any model of k-ekpyrosis where $P_{,X} > 0$ and the effective potential is positive semi-definite. While this form of superluminality may not give rise to causal paradoxes or otherwise violate fundamental physical principles, one must take extra care in such theories to account for the expanded causal structure of spacetime, the stricter requirements for well-posedness of initial conditions, and the possibility that UV completion may require non-standard approaches. 

It also remains unknown whether k-ekpyrosis models can perform as robustly as canonical theories in the non-linear regime where the universe is highly inhomogeneous, anisotropic, and spatially curved. The introduction of inhomogeneities would lead to both microscopic effects (e.g., through local gradient interactions) and macroscopic effects (in the sense that different regions of the universe may follow distinct trajectories) that were not accounted for in this work. The simultaneous presence of curvature and large anisotropies could also lead to chaotic mixmaster behavior and allow for new cosmological attractors that would not have been captured by the present analysis \cite{kasner, robustness}. Finally, it is not known whether the techniques employed to generate red-tilted power spectra of density perturbations in potential-driven models \cite{roman} would work equally well for k-ekpyrosis. Given these uncertainties and the previously outlined complications, the k-ekpyrosis scenario appears more challenging to implement than its potential-driven counterpart, but the simplicity and robustness of the high-$\epsilon$ power-law models in particular may make them a useful point of comparison in future analyses. 	
	
\section{Acknowledgements}
I wish to thank Paul Steinhardt for providing guidance throughout the study and writing process and to Anna Ijjas for her thoughtful suggestions and feedback on the manuscript. I am also grateful to Nima Arkani-Hamed and Giorgi Tukhashvili for their helpful discussions and comments. This work was supported in part by the DOE grant number DEFG02-91ER40671 and by the Simons Foundation grant number 654561.

\appendix
\section{Derivation of scaling solutions in exponential DBI models}\label{app:scaling_dbi}
To identify scaling solutions satisfying $\dot{\epsilon} = 0$, we can use Eq. (\ref{epsilon_dbi}) to re-express
\begin{equation} \label{wdot}
	\dot{\epsilon} = \frac{3\partial_t[(2\epsilon/3 - 1)^2]}{4(2\epsilon/3 - 1)} = \frac{3\partial_t(1-2fX)}{4(2\epsilon/3 - 1)} = -3\frac{f_\varphi \dot{\varphi}^3/2 + f\dot{\varphi}\ddot{\varphi}}{2(2\epsilon/3 - 1)}.
\end{equation}

We now use equation of motion (\ref{eom_dbi}), the warp factor (\ref{f_exp}), and the relations
\begin{align} 
	f(\varphi) &= \frac{\rho - P}{\rho P}, \\
	X &= \frac{P(\rho + P)}{2\rho}
\end{align}
[derived from Eqs. (\ref{p_dbi}) and (\ref{rho_dbi})] to write
\begin{widetext}
\begin{align} \label{exp_fp}
	\dot{\epsilon} &= \frac{-3f\dot{\varphi}^3}{2(2\epsilon/3-1)} \left[ \frac{1}{m} - \frac{3}{2}\frac{f_\varphi}{f} + \frac{f_\varphi}{f^2\dot{\varphi}^2} - 3H\dot{\varphi}^{-1}(1-2fX) - \frac{f_\varphi}{f^2\dot{\varphi}^2}(1-2fX)^{3/2} \right] \\ \label{fp_intermediate}
	&= \frac{-3f\dot{\varphi}^3}{2(2\epsilon/3-1)} \left[\frac{1}{m} - \frac{3}{m} + \frac{9}{2m\epsilon(3-\epsilon)} +\sqrt{3\rho}\,sign(\dot{\varphi}) \sqrt{\frac{(2\epsilon/3-1)^3}{P+\rho}}- \frac{9(2\epsilon/3-1)^3}{2m\epsilon(3-\epsilon)} \right] \\
	&= \frac{-3f\dot{\varphi}^3}{2m(2\epsilon/3-1)} \left[ -4 + \frac{4\epsilon}{3} + \frac{3}{\epsilon} +\sqrt{3}\,m\,sign(\dot{\varphi}) \sqrt{\frac{(2\epsilon-3)^3}{18\epsilon}} \right] \\ \label{fp}
	&= \frac{-3f\dot{\varphi}^3}{2m(2\epsilon/3-1)} \sqrt{\frac{(2\epsilon-3)^3}{18\epsilon}} \left[ 2\sqrt{1 - \frac{3}{2\epsilon}}+\sqrt{3}\,m\,sign(\dot{\varphi}) \right].
\end{align}
\end{widetext}
Note that in the second and fourth lines we assumed that $\epsilon > 3/2$, which necessarily holds true for models with negative $f(\varphi)$, and we have also assumed $H < 0$ throughout.

Because the factor outside the brackets is always nonzero (except when $\dot{\varphi} = 0$, in which limit we find $\epsilon = 3$), we can set the factor within the brackets equal to zero and conclude that a second scaling solution 
\begin{equation}
	\epsilon_{nc} = \frac{6}{4-3m^2}
\end{equation}
exists if $2/3 < m^2 < 4/3$ and $sign(\dot{\varphi}) = -1$ [or, more generally, $sign(H\dot{\varphi}) = 1$]. 

We now proceed to test the stability of these scaling solutions with a dynamical systems analysis. 
Up to positive overall coefficients, we may write

\begin{equation}
	\epsilon' \propto -\epsilon(\epsilon-3)(2\epsilon/3-1)  \left[ 2sign(\dot{\varphi})\sqrt{1 - \frac{3}{2\epsilon}}+\sqrt{3}\,m \right],
\end{equation}
where the prime denotes differentiation with respect to log$(a)$.

We can evaluate the stability of scaling solutions by evaluating the sign of $\frac{d\epsilon'}{d\epsilon}$ at the corresponding fixed-point values of $\epsilon$. If the sign is positive, then $\epsilon'$ is positive above the fixed point and negative below; this corresponds to stability in a contracting universe [since increasing log$(a)$ corresponds to decreasing time]. Analogously, if the sign is negative, the scaling solution is a repeller. Using this technique, we find the following:
\begin{itemize}
	\item For trajectories with $sign(\dot{\varphi}) > 0$, at the only fixed point $\epsilon = 3$, we have $\frac{d\epsilon'}{d\epsilon} < 0$ and hence the scaling solution is a repeller.
	\item For trajectories with $sign(\dot{\varphi}) < 0$ and $2/3 < m^2 < 4/3$, we have two fixed points to consider. At $\epsilon = 3$, we have $\frac{d\epsilon'}{d\epsilon} < 0$ and the scaling solution is once again a repeller. At $\epsilon = \epsilon_{nc}$, we have $\frac{d\epsilon'}{d\epsilon} > 0$ and hence the high-$\epsilon$ scaling solution is an attractor.
\end{itemize}

\section{Existence of DBI scaling solutions during subdominance}\label{app:scaling_sub}
To account for curvature, anisotropy, or other forms of stress-energy in the universe, we must retract our substitution of $|H| = \sqrt{\rho_\varphi/3}$ in Eq. (\ref{fp_intermediate}), which modifies the fixed point condition (\ref{fp}) to read
\begin{multline}
		\dot{\epsilon} = \frac{-3f\dot{\varphi}^3}{2m(2\epsilon_\varphi/3-1)} \sqrt{\frac{(2\epsilon_\varphi-3)^3}{18\epsilon_\varphi}}\\ \times \left[ 2\sqrt{1 - \frac{3}{2\epsilon_\varphi}}+\sqrt{3}\,m \frac{|H|}{\sqrt{\rho_\varphi/3}}\,sign(\dot{\varphi}) \right].
\end{multline}
The correction $|H|/\sqrt{\rho_\varphi/3} > 1$ amplifies the magnitude of the second term for any added form of positive energy density or negative curvature. If the correction is large enough (i.e., if the DBI energy density $\rho_\varphi$ is sufficiently small compared to the total energy density $3H^2$), then there will be no value of the DBI equation of state $\epsilon_\varphi$ that solves
\begin{equation}\label{scaling_sub}
	 2\sqrt{1 - \frac{3}{2\epsilon_\varphi}}-\sqrt{3}\,m \frac{|H|}{\sqrt{\rho_\varphi/3}} = 0
\end{equation}
and produces the $\epsilon_{nc}$ scaling solution for negative-velocity trajectories. Quantitatively, the restriction $\epsilon_\varphi \geq 3$ means that a solution to Eq. (\ref{scaling_sub}) only exists at a time $t$ when
\begin{equation}\label{omega_bound}
	\Omega_\varphi(t) > \frac{3m^2}{4}\cdot \frac{2\epsilon_\varphi(t)}{2\epsilon_\varphi(t) - 3};
\end{equation}
which automatically rules out the existence of the second scaling solution when $\Omega_\varphi < \frac{3m^2}{4}$.
	
\bibliographystyle{apsrev4-2}
%

\end{document}